\documentclass[twocolumn,aps,prl,longbibliography]{revtex4-2}
\usepackage{amsmath}
\usepackage{amssymb}
\usepackage{graphicx}
\usepackage[utf8]{inputenc}
\usepackage[T1]{fontenc}
\usepackage{color}

\renewcommand{\i}{\mathrm{i}}
\usepackage{braket}

\begin{document}
\title{Ultrastrong capacitive coupling of flux qubits}

\author{María Hita-Pérez}
\affiliation{Institute of Fundamental Physics IFF-CSIC, Calle Serrano 113b, 28006 Madrid, Spain}

\author{Gabriel Jaumà}
\affiliation{Institute of Fundamental Physics IFF-CSIC, Calle Serrano 113b, 28006 Madrid, Spain}

\author{Manuel Pino}
\affiliation{Institute of Fundamental Physics IFF-CSIC, Calle Serrano 113b, 28006 Madrid, Spain}

\author{Juan José García-Ripoll}
\affiliation{Institute of Fundamental Physics IFF-CSIC, Calle Serrano 113b, 28006 Madrid, Spain}

\begin{abstract}
  A flux qubit can interact strongly when it is capacitively coupled to other circuit elements. This interaction can be separated in two parts, one acting on the qubit subspaces and one in which excited states mediate the interaction. The first term dominates the interaction between the flux qubit and an LC-resonator, leading to ultrastrong couplings of the form $\sigma^y(a+a^\dagger),$ which complement the inductive $\sigma^xi(a^\dagger-a)$ coupling. However, when coupling two flux qubits capacitively, all terms need to be taken into account, leading to complex non-stoquastic ultrastrong interaction of the $\sigma^y\sigma^y$, $\sigma^z\sigma^z$ and $\sigma^x\sigma^x$ type. Our theory explains all these interactions, describing them in terms of general circuit properties---coupling capacitances, qubit gaps, inductive, Josephson and capactive energies---, that apply to a wide variety of circuits and flux qubit designs.
\end{abstract}
\maketitle

\paragraph{Introduction.--}
Flux qubits are superconducting loops with one or several Josephson junctions that, when the qubit is threaded by a magnetic flux, create a frustrated inductive energy landscape. The qubit's low-energy space is built from quantum superpositions of persistent current states with opposite directions\ \cite{orlando1999,mooij1999,chiorescu2003,peltonen2018}. Flux qubits exhibit strong magnetic interactions and large anharmonicities while retaining good coherence times\ \cite{You2007,bylander2011,braumuller2020}.  These are useful properties to implement fast qubit gates\ \cite{Liu2006,saito2009}, perform quantum annealing\ \cite{kadowaki1998,Bunyk2014,Boixo2014,weber2017,Hauke2020} or to simulate strongly coupled quantum systems\ \cite{niemczyk2010circuit,forn2010,peropadre2010,kakuyanagi2016,pino2015,forn2017,yoshihara2017,bernardis2018,zhu2011,pino2018,pino2020}.

A flux qubit is usually described by two observables: the tunneling between persistent current states $\sigma^z$, and the qubit's magnetic dipole moment $\sigma^x.$ The former accounts for the energy splitting $\Delta$ between current superpositions due to tunneling at the symmetry point $H=\Delta\sigma^z/2$\ \cite{mooij1999}. The dipole moment appears in the inductive coupling to microwave photons $g\sigma^xi(a^\dagger-a)$ and to other flux qubits $J_{xx}\sigma^x_1\sigma^x_2$. An open question is how to escape the narrow framework provided by these interactions, allowing flux qubits to simulate non-stoquastic Hamiltonians, the ones that exhibit a sign problem when solving them via Quantum Monte-Carlo\ \cite{bravyi2006,susa2017,albash2019role,hormozi2017, ozfidan2020} and that enable universal adiabatic quantum computation\ \cite{albash2018adiabatic,kitaev2002classical,kempe20033,kempe2006complexity,oliveira2005complexity}.

There is still no complete satisfactory framework to implement different flux-qubit couplings. One may obtain a ZZ interaction by converting the main qubit junction into a SQUID, and coupling those SQUIDs to each other inductively\ \cite{orlando1999}. However, the resulting interaction is weak and decreases monotonically with the gap, making it unsuitable for general quantum simulation and quantum annealing\ \cite{kerman2019}.  A similar coupling for qubits displaying non-trivial topological effects\ \cite{friedman2002,bell2016spectroscopic,kalashnikov2020} could give the desired interaction\ \cite{kerman2019}, but those qubits may suffer from an enhanced susceptibility to charge noise\ \cite{orlando1999,Chirolli206}. The most promising approach so far is the capacitive coupling between flux qubits. Experiments with flux qubits\ \cite{satoh2015,ozfidan2020} have demonstrated capacitve interactions along more than one direction\ \cite{consani2020effective},  but the coupling strength seems to be limited and there is no analytical framework to understand the range of available interactions.

\begin{figure}[b]
\includegraphics[width=1.\columnwidth]{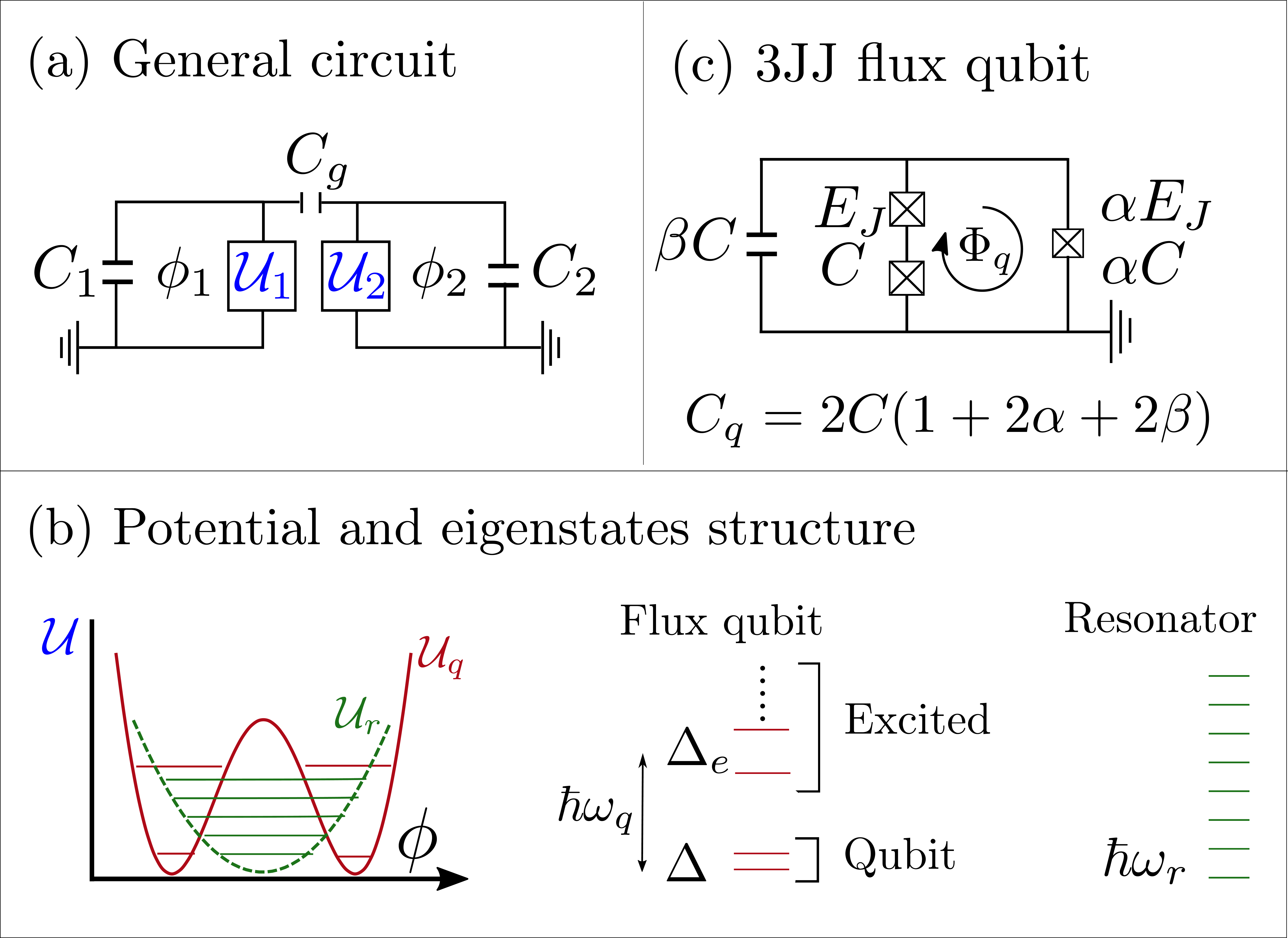}
\caption{ (a) Two superconducting circuits, described by flux variables $\phi_i$, coupled by a capacitance $C_g.$  (b) The first circuit is a flux qubit and the second is either an identical qubit or an LC-resonator.  The flux qubit, operated at full frustration, is described by an inductive potential $\mathcal{U}_q$ with two identical wells (solid line). The relevant energy scales are the qubit $\Delta$, the energy differences between states in the same well $\hbar\omega_q$ and, for deep potentials, the splitting $\Delta_e$ between excited states. The resonator has a quadratic potential $\mathcal{U}_r$ (dashed) and a harmonic spectrum with equispaced energies, separated by $\hbar\omega_r$. (c) Our simulations use three Josephson junction flux qubits, with a central junction $\alpha$ times smaller, a possible shunting capacitance $\beta$, and a magnetic flux $\Phi\simeq \Phi_0/2$. The charging and Josephson energies of the large junctions are denoted by $E_J$ and $E_C= e^2/(2C)$}.
\label{Fig1}
\end{figure}

In this work we introduce an analytical model for the capacitive coupling of flux qubits to other superconducting circuit elements [cf. Fig.\ \ref{Fig1}], complemented by a non-perturbative numerical treatment. We obtain the shape and scaling of the interaction between a qubit and a microwave resonator, and also between two flux qubits. We provide evidence of ultrastrong qubit-resonator interaction mediated by the electric dipole moment $\sigma^y$, thereby extending the family of ultrastrong inductive couplings\ \cite{niemczyk2010circuit,forn2010,yoshihara2017}, which are mediated by $\sigma^x$ terms. For two capacitively coupled qubits, we explain the appearance of complex interactions along multiple directions, YY, ZZ and XX. This is a rich landscape of spin Hamiltonians for quantum simulation, quantum annealing and quantum computation, which exceeds the simple picture from spectroscopic characterizations\ \cite{ozfidan2020} and complements earlier numerical studies\ \cite{consani2020effective} with scalings based on qubit's design parameters and a modelization of interactions mediated by excited states.

\paragraph{Model.--} We study on equal footing both the qubit-qubit and qubit-resonator capacitive couplings, as shown in Fig.\ \ref{Fig1}. Each element, qubit or resonator, is represented by a flux $\phi$ and a charge $q$ operator, and the type of circuit is determined by the inductive potential $\mathcal{U}$. The Hamiltonian $H = H^{(0)}+\epsilon V$ splits into bare circuits and interaction\ \cite{suppl}
\begin{align}
    H^{(0)}&=\sum_{i=1,2} \frac{q_i^2}{2C_i}+\mathcal{U}_i(\phi_i),\\
    V &= -\frac{q_1q_2}{\sqrt{C_1C_2}}- \sqrt{\frac{C_2}{C_1}}\ \frac{q_1^2}{C_1}- \sqrt{\frac{C_1}{C_2}}\ \frac{q_2^2}{C_2}.\label{eq:coup}
\end{align}
$C_{1,2}$ are the capacitances of bare circuits, $C_g$ is the coupling capacitance and $\overline{C}_\text{od} = \left[C_1C_2+(C_1+C_2)C_g\right]/C_g$ is the off-diagonal of the inverse capacitance matrix. The strength of the coupling is controlled by $\epsilon=\sqrt{C_1C_2}/\overline{C}_\text{od},$ with $\epsilon\sim \mathcal{O}(C_g)$ for $C_g/C_{1,2}\ll 1.$

The first circuit is always a flux qubit $C_1= C_q$. The second circuit will be either an identical qubit $C_2=C_1$, or a microwave resonator $C_2=C_r$. Without loss of generality, we study three Josephson Junction flux qubits (3JJQ) [cf. Fig.\ \ref{Fig1}(c)], operating at full frustration, with half a flux quantum $\Phi_q=\Phi_0/2$ in the loop. In this situation, this or any other similar qubit will exhibit an inductive potential $\mathcal{U}_{1,2}$ with local minima at $\phi = \pm \varphi_*(\Phi_0/2\pi)$. Each minima is associated to one persistent current state and a local excitation energy $\hbar\omega_q$. We present results in terms of the 3JJQ's relative coupling $\gamma=C_g/C$. For the resonator we use a quadratic inductive potential $\mathcal{U}_2 = \phi_2^2/(2L_r)$ with resonator frequency $\omega_r=1/\sqrt{L_rC_r}.$ 

\paragraph{Methods.--} We model the whole system as a an effective qubit-resonator or qubit-qubit Hamiltonian, using a Schrieffer-Wolff transformation $U(\epsilon)$\ \cite{bravyi2011} that maps the eigenspaces of $H$ to the eigenspaces of the bare model $H^{(0)}$. In our analytical treatment, we start from a projector $P_0$ onto a low-energy subspace---e.g. the 4-dimensional space of two qubits, or a tensor product of a qubit and resonator spaces---and  develop the effective Hamiltonian as a perturbative series
\begin{equation}
 H_\text{eff} = P_0H^{(0)}P_0 +\sum_{n=1} \epsilon^n \mathcal{M}_n.\label{eq:series}
\end{equation}
The first order term is the projection of the interaction onto the qubit subspace $\mathcal{M}_1=P_0^\dagger V P_0$, while $\mathcal{M}_{i\geq 2}$ describe interactions mediated by virtual transitions\ \cite{suppl}.

Numerically, we could imitate this procedure\ \cite{consani2020effective}, but instead we sum the series to all orders\ \cite{hita2021,bravyi2011}, as $H_\text{eff}=P_0UHU^\dagger P_0$. The unitary transformation $U P = P_0 U$ is derived from the projector $P$ and $P_0$, onto the numerically exact eigenstates of $H$ and $H^{(0)}$. We then expand the effective model $H_\text{eff}$ using Pauli and Fock operators. This allows us to compare the effective Hamiltonian to the predictions from perturbation theory, validating the type and scaling of the coupling terms.

\paragraph{First order capacitive interaction.--} The interaction $V$ has two terms that renormalize the bare circuits, and only one that entangles their dynamics $V_{c} = - q_1q_2/\sqrt{C_1C_2}.$ To develop the first order correction $\mathcal{M}_1$ we must express the charges $q_i$ in the qubit and resonator basis. For the resonator $q_2=\sqrt{\hbar/2Z}(a^\dagger + a)$ exactly, in terms of Fock operators $\{a,a^\dagger\}$ and the resonator impedance $Z$. For the flux qubit $q_1$ we assume that the renormalized Hamiltonian is anharmonic and $H_1\simeq\hbar\Delta\sigma_1^z/2$. We derive the voltage operator projected onto the qubit subspace $\mathcal{V}_1$ as the derivative of the flux $\mathcal{V}_1\simeq i[H_1,\phi_1]/\hbar=(\Phi_0/2\pi)\frac{\varphi_\star\Delta}{\hbar}\sigma^y_1,$ approximating the flux operator as the flux jump between qubit states $\phi_1= (\Phi_0/2\pi)\varphi_\star \sigma^x.$ Since  $\mathcal{V}_1 = q_1/\overline{C}_1$ for the bare renormalized qubit, the projected charge operator is $q_1\simeq \Phi_0\overline{C}_1\Delta \varphi_\star\sigma_1^y/h .$

Using this method, we obtain the first order effective interaction between two qubits $H^{(1)}_{qq} = g^{(1)}_{qq} \sigma_1^y \sigma^y_2$ with
\begin{align}
    \frac{g^{(1)}_{qq}}{\Delta} &= \frac{\overline{C}_q  \varphi_\star^2}{\overline{C}_\text{od}}  \frac{\Delta}{8 E^{q}_C}. \label{eq:qqfirst}
\end{align}
and the interaction between a qubit and a resonator $H^{(1)}_{qr} = i g^{(1)}_{qr} \sigma_1^y (b^\dagger- b)$ with
\begin{align}
    \frac{g^{(1)}_{qr}}{\Delta} &= \frac{\overline{C}_q}{\overline{C}_\text{od}}\frac{\varphi^\star}{2}\sqrt{\frac{1}{2\pi G_0\mathcal{Z}}}.\label{eq:qrfirst}
\end{align}
Everything depends on the qubit's renormalized gap $\Delta$, the resonator impedance $Z$, the conductance quantum $G_0$ and the qubit's charging energy $E_C^q= e^2/(2\overline{C}_1)$.

This treatment neglects higher order terms in the perturbation series\ \eqref{eq:series}, generated by matrix elements of the qubit's charge operator $(\openone-P_0)q_1 P_0$ connecting qubit states with excited states delocalized among the inductive potential wells. Those elements grow as $q_1\sim\sqrt{\omega_q}$, requiring us to analyze their effect in a case-by-case basis.

 \begin{figure}[t!]
\begin{centering}
\includegraphics[width=1.0\columnwidth]{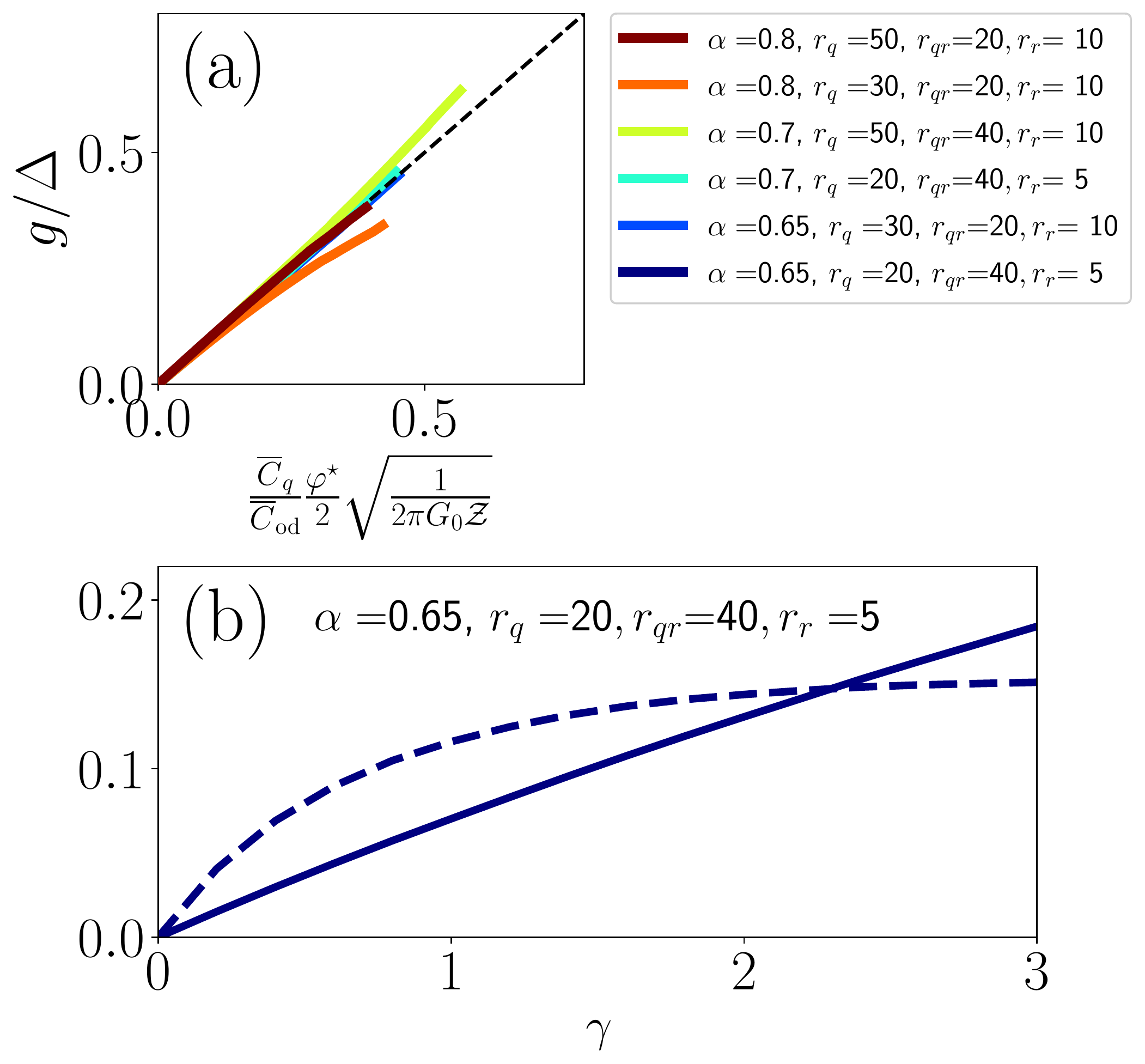}
\end{centering}
\caption{ A 3JJQ capacitively couple to a LC-resonator. (a) Coupling divided by qubit gap $\Delta$ as a function of the first order corrections (the dashed line is the theoretical prediction for those corrections). The design parameters for all the panels are qubit large junction energy ratio $r_q=E_J/E_C$, energy ratio for the resonator $r_r=E^r_J/E^r_C$ and ratio of resonator and qubit Josephson energies $r_{qr}=E_J/E_J^r.$ The Josephson energy for the resonator as a function of its inductance $L_r$ is $E_{J}^r = \frac{1}{L_r}\left(\frac{\hbar}{2e}\right)^2.$ (b) Coupling divided by qubit gap (solid line) and resonator energy (dashed) as a function of the ratio $\gamma=C_g/C$ between shared and qubit's large junction capacitances. 
}\label{Fig_qr}
\end{figure}

\paragraph{Strong qubit-resonator coupling.}
Let us discuss the capacitive coupling of a 3JJQ [c.f. Fig.\ \ref{Fig1}(c)] with an LC-resonator. The second order corrections to the capacitive coupling involve a simultaneous excitation of the qubit and the resonator, which acquire an energy $\hbar\omega_q$ due to leaving the qubit space and $\hbar\omega_r$ due to the creation or annihilation of a photon. While the amplitude of these processes in $\epsilon V_c$ grows as $\hbar\sqrt{\omega_q\omega_r}$, the resonator cannot easily absorb the energy $\hbar\omega_q.$ Thus, higher-order corrections $\mathcal{M}_{i\geq 2}$ only renormalize the qubit's self-energy and the capacitive coupling is fully captured by Eq.\ \eqref{eq:qrfirst}.

We confirm this hypothesis with the full Schrieffer-Wolff transformation of the qubit-resonator circuit model~\eqref{eq:coup}. In terms of qubit and photon operators, it takes the form $H_\text{eff}\simeq \hbar\Delta\sigma^z/2 + \hbar\omega a^\dagger a + g \sigma^y (a^\dagger + a)$ for moderate numbers of photons. As shown in Fig.\ \ref{Fig_qr}(a), all the coupling constants from different designs of qubit and resonator collapse into the first order correction derived analytically\ \eqref{eq:qrfirst}, up to coupling strengths $g/\Delta\approx 1$ beyond the perturbative limit. Since our model only demands a qubit anharmonic spectrum, we conclude that\ \eqref{eq:qrfirst} is a general theory for the capacitive interaction between flux qubits and microwave resonators.

We also have evidence that the capacitive coupling can enter the ultrastrong coupling regime, where $g/\omega_r\sim g/\Delta \geq 12\%$. Unlike the inductive case\ \cite{niemczyk2010circuit,forn2010,yoshihara2017}, exploring the actual designs where this happens is complicated, because one has to consider the renormalization of the qubit's gap, while the resonator remains more or less unperturbed. Fig.\ \ref{Fig_qr}(c) shows evidence of this regime using a qubit with an intermediate ratio of qubit Josephson and charging energies $r_q=E_J/E_C=20$ which slows down the gap renormalization. Doing so, we achieve the ultra-strong coupling regime $g\approx 0.15\Delta$ at zero detuning $\Delta=\omega_r$. We expect future simulations will reveal more favorable situations, by tuning both the qubit's and the resonator's impedances.

\begin{figure}[t]
\begin{centering}
\includegraphics[width=\columnwidth]{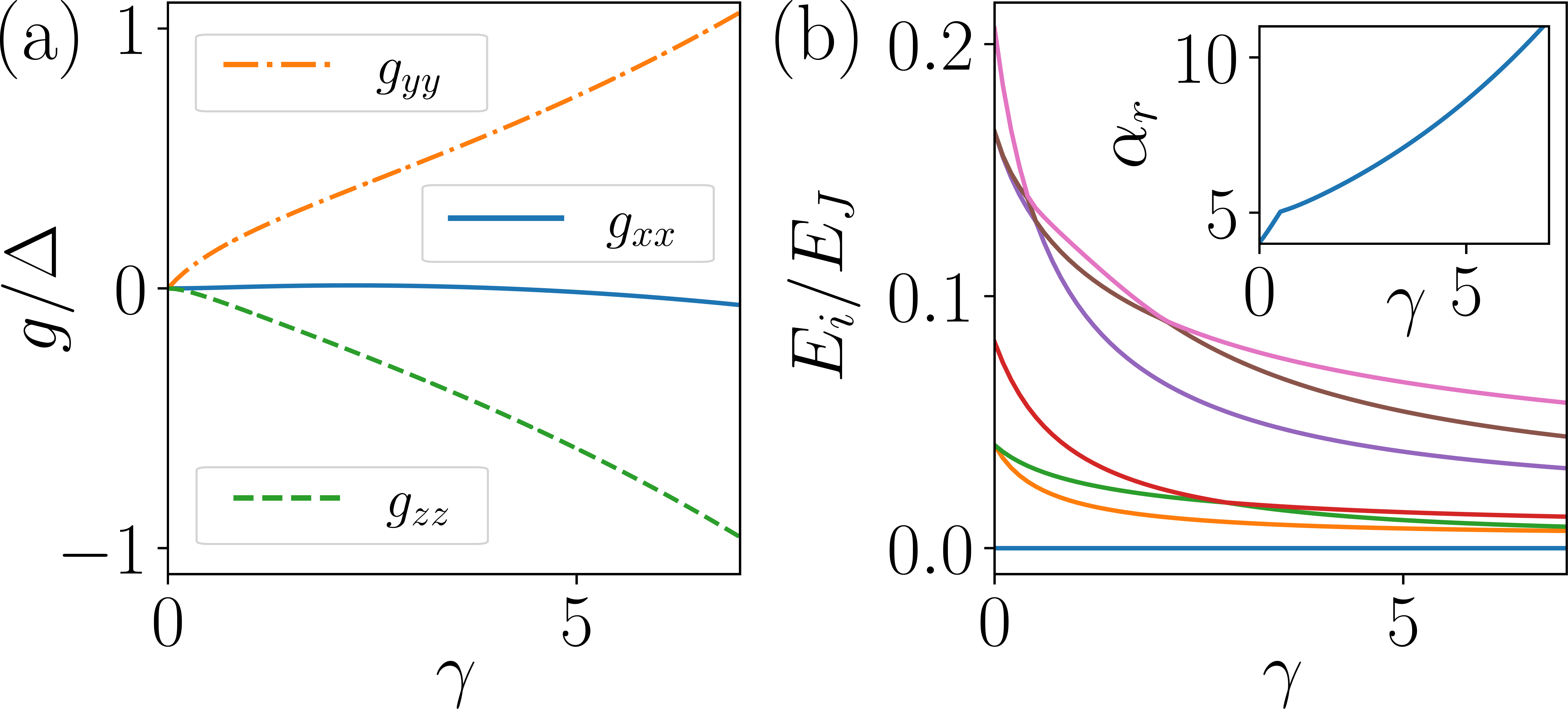}
\end{centering}
\caption{ (a) Coupling strength in gap units and (b) low energy spectrum for two equal 3JJQ capacitively coupled, both having $\alpha=0.65$ and ratio $E_J/E_C=50,$ as a function of the ratio $\gamma=C_g/C$ between shared and qubit's large junction capacitances. The $E_J,E_C$ are Josephson and charging of the large junction qubit. The inset of (b) shows the relative anharmonicity of each qubit approximated from the full spectrum as $\alpha_r = (E_3-E_{0}-\Delta)/\Delta.$
}\label{Fig3}
\end{figure}

\paragraph{Qubit-Qubit coupling.--}
Let us now discuss the capacitive interaction between two flux qubits. We begin by presenting the numerically exact Schrieffer-Wolff transformation for two 3JJQ's. In Fig.\ \ref{Fig3}(a) we plot the interaction coefficients that result from expanding $H_\text{eff}$ in the basis of Pauli matrices. The numerical model clearly shows the first order terms associated to the $\sigma^y$ charge dipole operator in Eq.\ \eqref{eq:qqfirst}. However, the flux qubits acquire also a comparable ZZ interaction that enables the tunneling of current states, and we also find a residual inductive XX coupling that explodes once $\gamma =C_g/C \gg 1$. Note while the capacitive term induces a renormalization of the qubit's gap $\Delta$, the qubit nature is preserved by an improvement in the qubit's relative anharmonicity $\alpha_r$ [c.f. Fig.\ \ref{Fig3}(b)]. Only at very large $\gamma,$ the form of the interactions approaches $-g(\tilde\sigma^+_1\tilde\sigma^+_2+\tilde\sigma^-_1\tilde\sigma^-_2),$ with ladder operators in the persistent current base $\tilde\sigma^{\pm}_i=\tilde\sigma_i^z\pm i\tilde\sigma^y_i$. In this limit, the coupling produces a large mass in the direction $\phi_1-\phi_2$ of the two qubit system and strongly suppress transitions of the form $\tilde\sigma^+_1\tilde\sigma^-_2+\tilde\sigma^-_1\tilde\sigma^+_2$ \cite{levitov2001}.

\begin{figure}[t!]
\begin{centering}
\includegraphics[width=1.\columnwidth]{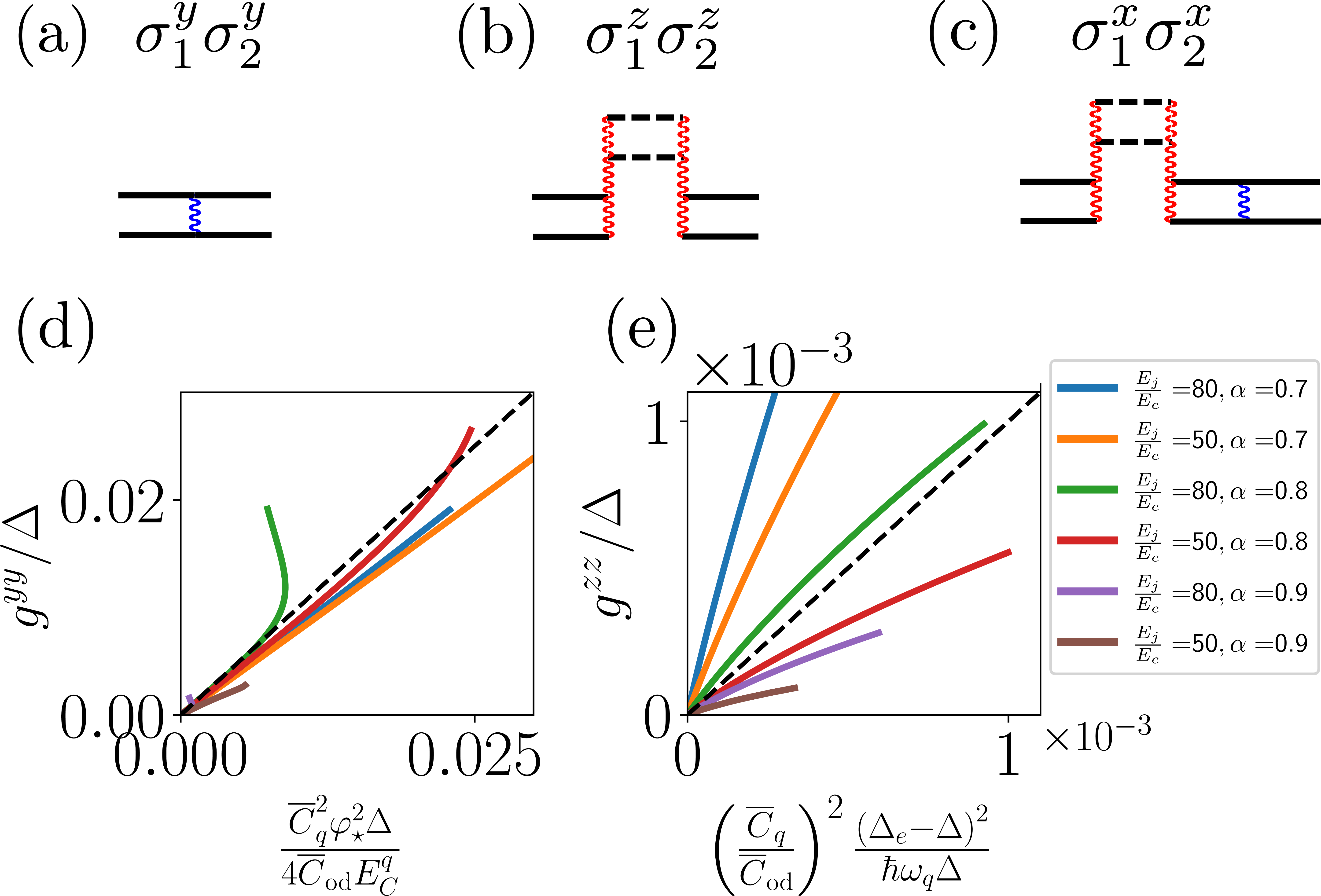}
\end{centering}
\vspace{0.1cm}
\caption{ Diagrams for first (a), second (b) and third (c) order corrections to the capacitive coupling of two 3JJQ. Qubit and excited states are represented by solid and dashed lines, respectively.  Wiggle lines are qubit interactions, which can be diagonal (shorter blue line) and off-diagonal (longer red) in the qubit subspace. First order corrections are the projection of charge-charge operators to the qubit subspace. It gives a $YY$ type of coupling.  Second order produces a $ZZ$ type of coupling. We only show one of the possible diagrams for third order corrections. It can be seen to give $XX$ coupling using the lower order diagrams. Data collapse of the (e) first and (d) second order corrections for several qubit design parameters.
}\label{Fig4}
\end{figure}

We interpret these results using the perturbation theory\ \eqref{eq:series} processes $\mathcal{M}_1$, $\mathcal{M}_2$ and $\mathcal{M}_3$, sketched in Figs.\ \ref{Fig4}(a-c). The horizontal lines denote qubit (solid) and excited states (dashed). These are connected by interactions (wiggly lines), which can be qubit terms $P_0VP_0$ (blue) or connect to excitations $P_0V(\openone-P_0)$ and $(\openone-P_0)VP_0$ (red). To first order, the capacitor produces YY terms in the qubit space. To second order, the operator $V_c$ enables virtual transitions where both circuits momentarily excite, acquiring an energy $\sim 2\times \hbar\omega_q$. Despite the large energy difference, these processes are assisted by matrix elements in $V_c$ that grow as $\omega_q$, and cannot be neglected. We estimate the second order term $H_{qq}^{(2)} =g^{(2)}_{qq} \sigma_1^z\sigma_2^z$ with
\begin{align}
    g^{(2)}_{qq}=\frac{\overline{C}_q}{\overline{C}_\text{od}}\frac{(\Delta_e-\Delta)^2}{\hbar \omega_q},\label{eq:zz}
\end{align}
where $\Delta_e$ is the approximate splitting between excited states due to tunneling [See\ \cite{suppl} and Fig.\ \ref{Fig1}a]. Finally, we can also deduce that third order terms create an XX coupling. This coupling becomes dominant when the renormalization of the qubit's capacitance enables phase-slip transitions between the qubit's unit cells---as opposed to the tunneling between current states enabled by $\Delta$---. However, this mechanism leads to an enhancement of charge noise\ \cite{orlando1999,Chirolli206}. We can stay far from this regime by limiting $\gamma\sim O(1)$ and choosing $\alpha\ll 0.9$, away from the single qubit phase slip regime\ \cite{orlando1999}. This way we can still recover an XX interaction by coupling the qubits inductively, which in combination with YY an ZZ interactions gives a fully non-stoquastic spin model.

As in the resonator case, we compare our predictions\ \eqref{eq:qqfirst} and\ \eqref{eq:zz} to the numerically exact couplings, for different qubit designs. Fig.\ \ref{Fig3}(d) shows how the $g^{yy}$ coupling collapses to the theoretical prediction\ \eqref{eq:qqfirst} only for small enough couplings $g/\Delta<0.05$. Similarly, Fig.\ \ref{Fig3}(e) shows that $g^{zz}$ follows the expected scaling, up to factors $\mathcal{O}(1)$. Perturbation theory thus captures the overall dependency of the couplings on the qubit's parameters, but fails to estimate the non-perturbative corrections that account for the full interaction. This contrasts with the qubit-resonator model, and highlights the relevance of qubit-qubit interactions mediated by excited states.

\paragraph{Conclusions.--}
We have presented a non-perturbative study of capacitive interactions between a flux qubit and other circuits. This study reveals that the flux qubit charge operator is the sum of two dipole moments: one connecting qubit states and one enabling transitions to higher energy excitations. The first term describes the coupling between a qubit and a resonator and supports ultrastrong qubit-photon interactions along directions orthogonal to the ones created by inductive terms. The second term combines with the first one to enable a rich family of non-stoquastic qubit-qubit couplings, including YY, ZZ and XX interactions.

The capacitive interactions combined with inductive ultrastrong qubit-photon interactions\ \cite{niemczyk2010circuit,forn2010,forn2017,yoshihara2017} open new regimes in light-matter and light-mediated interactions, such as the ultrastrong coupling limit of the Jaynes-Cummings model\ \cite{huang2020}, new regimes of the spin-boson model\ \cite{forn2017} with two transverse couplings, and new models of coherent and dissipative interactions mediated by photon exchange, beyond those in Ref.\ \cite{gonzalez-tudela2011}.

Regarding quantum simulation, our study confirms the idea that flux qubits exhibit rich families of non-stoquastic interactions. These may appear combined, as in the two-qubit model, or they may be pure YY interactions, if we use resonators to mediate the coupling\ \cite{kurcz2014,pino2018}. From a fundamental point of view, it would be interesting to explore the long-range interactions at the hardware level, without embeddings\ \cite{Venturelli2015}. Indeed, capacitively coupled flux qubits in 2D geometries support YY interactions that only decay logarithmically up to a length $\xi=\sqrt{C_g/C_q}$ and then exponentially fast\ \cite{ortuno2015}. As in classical spin-glasses, such long-range interactions will produce hard to solve quantum models\ \cite{fernandez2018,fernandez2019, Marshall2016,pino2012,pino2020_2}.

Finally, recent works\ \cite{ciani2020,halverson2020efficient} suggest that the \textit{ground state properties} of many superconducting circuits, including capacitively coupled flux qubits, could be efficiently simulable in the semiclassical charge-flux representation. Our work opens a rigorous avenue to study these circuits and their effective models in a non-perturbative fashion. This will help us understand whether the ground states of circuits are in some sense trivial---e.g. low-energy states are classical spin configurations---, or whether the energy scales and types of interactions reveal other kinds of obstructions, different from the sign problem, that prevent the classical simulation of the device.

\begin{acknowledgments}
This work has been supported by European Commission FET-Open project AVaQus GA 899561 and CSIC Quantum Technologies Platform PTI-001. Financial support by Fundación General CISC (Programa Comfuturo) is acknowledged. The numerical computations have been performed in
the cluster Trueno of the CSIC.
\end{acknowledgments}

\bibliography{./annealing}

\newpage
\onecolumngrid
\appendix

\clearpage

\section{APPENDIX 1: Hamiltonian of two flux qubits coupled via a capacitor}\label{SM1}

We derive the Hamiltonian for a system of two flux qubits coupled via a capacitor. We assume two identical  qubits, with qubit capacitance $C_q$, which are coupled by a capacitor $C_g$ as in Fig.\ \ref{Fig1} of the main text. The Lagrangian in term of the capacitance matrix:
\begin{align}
    \mathcal{L} = \frac{1}{2} \vec{\dot{\phi}}\ \mathcal{C}\ \vec{\dot{\phi}}-\sum_{i=1,2}\mathcal{U}_i
\end{align}
with the vector of fluxes of each qubit $\vec{\phi}=(\phi_1,\phi_2)$ and capacitance matrix:
\begin{align}
    \mathcal{C} = \begin{bmatrix}
    C_q+C_g & -C_g  \\
    -C_g &  C_q+C_g
  \end{bmatrix}.
\end{align}
The Hamiltonian $H=\sum_{i=1,2}q_i\dot{\phi}_i-\mathcal{L}$ can be written using flux and its conjugate charge:
\begin{align}
    H = \sum_{i=1,2} \frac{q_i^2}{2\overline{C}_q} +\mathcal{U}(\phi_i) + \frac{q_1 q_2}{\overline{C}_\text{od}}.\label{eq:SM_H}
\end{align}
The inverse of capacitance matrix $\mathcal{C}^{-1}$ is used to obtain $1/\overline{C}_q=(\mathcal{C}^{-1})_{11}=(\mathcal{C}^{-1})_{22}$ and $1/\overline{C}_\text{od}=(\mathcal{C}^{-1})_{12}= (\mathcal{C}^{-1})_{21}$. We can obtain the following expression for the capacitances involved in Eq.\ \eqref{eq:SM_H} as:
\begin{align}
    \frac{1}{\overline{C}_q} &= \frac{1}{C_q} \frac{C_q+C_g}{2C_g+C_q} \\
    \frac{1}{\overline{C}_\text{od}} &= \frac{1}{C_q} \frac{C_g}{2C_g+C_q}
\end{align}
Taking into account that the renormalized qubit matrix is $\frac{1}{\overline{C}_q} = \frac{1}{C_q}-\frac{1}{\overline{C}_\text{od}},$ we express the Hamiltonian as a sum of the non-coupled Hamiltonian $H_0$ plus a perturbation $V$, as in Eq.\ \eqref{eq:coup} of the manuscript:
\begin{align}
    H &= H^{(0)} + \epsilon V \\
    H^{(0)} &= \sum_{i=1,2} \frac{q_i^2}{2 C_q} +\mathcal{U}(\phi_i)  \\
    V &= \frac{q_1 q_2}{C_q} - \sum_{i=1,2} \frac{q_i^2}{2C_q}
\end{align}
where $\epsilon=C_q/\overline{C}_\text{od}$. The second term in the previous Hamilton is a perturbation to the non-coupled system when $\epsilon\ll 1$. The Hamiltonian for the general case Eq.\ \eqref{eq:coup} with asymmetric capacitances $C_1,C_2$ can be obtained from the inverse of the capacitance matrix  $1/\overline{C}_1 = 1/C_1-C_2/(\overline{C}_\text{od} C_1)$ and $1/\overline{C}_2 = 1/C_2-C_1/(\overline{C}_\text{od}C_2)$.

\section{APPENDIX 2: Perturbation theory for the capacitive coupling of two flux qubits}\label{SM2}

We use the Schrieffer--Wolff transformation\ \cite{bravyi2011} and expand the effective Hamiltonian in series of the small parameter $\epsilon$, as in Eq.\ \eqref{eq:series} of the main part of the manuscript. The perturbation series up to third order can be expressed as:
\begin{align}
    H_\text{eff} = H^{(0)} + \epsilon \mathcal{M}_1 + \frac{\epsilon^2}{2} \mathcal{M}_2 + \frac{\epsilon^3}{2} \mathcal{M}_3
\end{align}
with matrix:
\begin{align}
\mathcal{M}_1 &= P_0 V P_0\\
\mathcal{M}_2 &= P_0 \hat{S}(V_\text{od}) P_0 \\
\mathcal{M}_3 &= P_0 \hat{V}_\text{od} \mathcal{L}\hat{V}_{d}(S) P_0
\end{align}
where the adjoint representation is $\hat{Y}(x) = [Y,X]$ and the operators $P_0, Q_0 $ project onto the qubit and excited subspaces of the unperturbed system. The notation $O_\text{od}$ is used for the non-diagonal part of an operator $O_\text{od}=P_0OQ_0+Q_0OP_0$ and $S = \mathcal{L}(V_\text{od}),$ being
$\mathcal{L}(O) = \sum_{ij} \left(O_\text{od}\right)_{ij} / E_{ij} \ket{j}\bra{i}$ (matrix elements are denoted by $O_{ij}=\braket{i|O|j}$ and energy differences $E_{ij}=E_i-E_j$).

We now write the explicit forms for the matrix involved in the computation of the effective Hamiltonian up to third order. We do so by employing Latin and Greek letters to denote unperturbed qubit and excited states, respectively:
\begin{align}
\mathcal{M}_1 &= \sum_{i,j} V_{ij}\ket{j}\bra{i}\label{S_eq:1st}\\
\mathcal{M}_2 &= \sum_{i,j,\alpha} \frac{V_{i\alpha}V_{\alpha j} }{E_{i\alpha}} \left(\ket{j}\bra{i}-\ket{i}\bra{j}\right). \label{eq:2nd}\\
\mathcal{M}_3 &= \left(\sum_{\alpha,j,\beta,k} \frac{V_{j \alpha}V_{\alpha \beta}V_{\beta k}}{E_{\alpha j}E_{\beta j}} -  \sum_{\alpha,i,j,k} \frac{V_{j i}V_{i \alpha}V_{\alpha k}}{E_{\alpha i}E_{\alpha j}}
\right) \left( \ket{k}\bra{j}-\ket{j}\bra{k}\right)\label{eq:SM_3rd}
\end{align}
the braket of the interactions in the qubit subspace is $V_{ij}=\braket{i|V|j},$ on the excited one $V_{\alpha \beta}=\braket{\alpha|V|\beta}$ and the off diagonal between qubit and excited states $V_{\alpha i}=\braket{\alpha|V|i}.$

As discussed in the main body of the work, the first order corrections Eq.\ref{S_eq:1st} are given by the projection of the perturbation to the qubit subspace. They produce a reonormalization of the gap mass and a coupling of the $YY$ type. The higher order corrections that couple the qubits involve operators that move the state of the superconducting circuit from qubit to excited states. We analyze in the following those corrections.

\section{APPENDIX 3: Second and third order corrections to qubit-qubit coupling.}\label{SM3}

We focus now on the second order correction for two three Josephson junctions qubits (3JJQ) coupled via a capacitor. We do not analyze contributions that renormalizes each qubit gap, only contributions that couple the two qubit system. The relevant diagram  is shown in panel in Fig.\ \ref{Fig3}(b) of the main manuscript, where the qubits visit high energy states due to the off-diagonal part of the interaction. We need to make strong simplifications on the qubit spectrum, as we explain in what follows, in order to treat analytically second order corrections. Each qubit is approximated by  the first fourth eigenstate of the uncoupled superconducting circuit. We use the notation $\ket{\pm_g} = (\ket{L_e}\pm\ket{R_e})$ for the qubit eigenstates and  $\ket{\pm_e} = (\ket{L_e}\pm\ket{R_e})$ for second and third excited states, being $\ket{L_g}$ and $\ket{L_e}$ the ground and first excited state inside the left well of the potential and similar notation for the right well $\ket{R_g},\ \ket{R_e}$, see Fig.\ \ref{Fig1}(b) of the main text for a picture of the spectrum.

The unperturbed projected Hamiltonian onto the subspace expanded by the first four states of each qubit is:
\begin{align}
    H^{(0)} =  \sum_{i=1,2} \frac{\Delta}{2} P_{0i}^\dagger\sigma_i^z P_{0i} + \sum_{i=1,2} Q_{0i}^\dagger\left(\hbar \omega + \frac{\Delta_e}{2}
    \sigma_i^z\right)Q_{0i}
\end{align}
where $P_{0i}$ and $Q_{0i}$ projects onto the unperturbed low and excited subspace of qubit $i$, respectively. We have extended the domain of Pauli matrices so that they act on ground $\sigma_i^z\ket{\pm_g}=\pm\ket{\pm_g}$ and excited states $\sigma_i^z\ket{\pm_e}=\pm\ket{\pm_e}.$ In order to obtain the second order term, we need to compute the non-diagonal energy between the low-energy and high energy sector. We recall that, in the case treated here, $V_\text{od}=P_0VQ_0+Q_0VP_0$, so that we can approximate the off-diagonal elements of the interaction as:
\begin{align}
    V_{i_g,\alpha_e} =\braket{i_g|q_1q_2|\alpha_e} =\hbar\omega_q \delta_{i\alpha}\label{eq:vof}
\end{align}
where $i,\alpha=1,2,3,4.$ The low energy sector of the coupled system is $\ket{1_{g}}=\ket{+_{g},+_{g}},$ $\ket{2_{g}}=\ket{+_{g},-_{g}},$ $\ket{3_{g}}=\ket{-_{g},+_{g}}$ and $\ket{4_{g}}=\ket{-_{g},-_{g}}$. Similarly,  we have for the excited states $\ket{1_e}=\ket{+_e,+_e},$ $\ket{2_{e}}=\ket{+_{e},-_{e}},\dots$. The effective harmonic frequency of each of the single-qubit potential wells is $\omega_q,$ see Fig.\ \ref{Fig1}(b) of the main text. We do not consider matrix process that only take one qubit outside of the low-energy subspace and bring it back. Those processes have an amplitude that scale as $ \sqrt{\hbar \omega_q \Delta},$ which is much smaller than the one which takes the two qubits outside of the qubit subspace in Eq.\ \eqref{eq:vof} (we are in the anharmonic limit $\Delta\ll\hbar \omega_q$). We then approximate the second order corrections as:
\begin{align}
    \mathcal{M}_2 = 2(\hbar \omega_q)^2 \left(
    \frac{\ket{+_g+_g}\bra{+_g+_g}}{\hbar\omega_q+2(\Delta_e-\Delta)}  +
    \frac{\ket{-_g-_g}\bra{-_g-_g}}{\hbar\omega_q-2(\Delta_e-\Delta)} +
    \frac{\ket{-_g+_g}\bra{-_g+_g}+\ket{+_g-_g}\bra{+_g-_g}}{\hbar\omega_q}
    \right)
\end{align}
Expanding the denominator up to second order in $\Delta_e-\Delta$ and using the identity $\ket{\pm}\bra{\pm}=(1\pm \sigma^z)/2,$ we found that:
\begin{align}
    H^{(2)} &= g^{(2)}_{qq} \sigma^z_1 \sigma_2^z.\\
    g^{(2)}_{qq} &= \frac{\epsilon^2}{2} \frac{(\Delta_e - \Delta)^2}{\hbar \omega_q}
\end{align}
In the case of two 3JJQs, we can express previous formula using the ratio between large Josephson junction and coupling capacitances $\gamma=C_g/C$, instead of the parameter in the perturbative expansion via $\epsilon=C_q/\overline{C}_\text{od}=\gamma/(1+2(\alpha+\gamma+\beta))$. The second order corrections are then:
\begin{align}
g^{(2)}_{qq} = \frac{\gamma^2}{8\hbar \omega_q} \left(\frac{\Delta_e-\Delta}{1+2\alpha+2\beta }\right)^2,
\end{align}
This is the formula that is employed to plot results in Fig.\ \ref{Fig4} of the main text. 

Using similar approximations as before, we can find that third order corrections induces a coupling of the type $\sigma_1^x\sigma_2^x$. To do so, we analyze the third order correction correspond to the second term inside the parenthesis at the right hand side of Eq.\ \eqref{eq:SM_3rd}. These corrections can be approximated as:
\begin{align}
g^{(3)}_{qq}\approx\sum_{\alpha,i,j,k} \frac{V_{j i}V_{i \alpha}V_{\alpha k}}{E_{\alpha i}E_{\alpha j}}
 =     \frac{\mathcal{M}_1}{\hbar \omega_q} \mathcal{M}_3
\end{align}
Taking into account the shape of the first two corrections, this terms would give a contribution to coupling $\sigma_1^x\sigma_2^x$ as explained in Fig.\ \ref{Fig3}. Although there are other third order corrections, the analysis performed here shows that the  coupling $XX$ dominates the third order. This is exactly the case in our numeric in Fig.\ \ref{Fig3} of the main part of the manuscript, where $g_{xx}$ depends on $\epsilon^3$ at small $\epsilon$.

\end{document}